\documentclass[12pt,preprint,aps,showpacs]{revtex4}
\usepackage{graphicx}
\usepackage{dcolumn}
\usepackage[normalem]{ulem}

\begin{document}

\title{First Observation of In-Medium Effects on Phase Space Distributions of Antikaons Measured
in Proton-Nucleus Collisions}

\author{
W.~Scheinast$^f$,  I.~B\"ottcher$^d$, M.~D\c{e}bowski$^{e,f}$,
F.~Dohrmann$^f$, A.~F\"orster$^{b,\dagger}$, E.~Grosse$^{f,g}$,
P.~Koczo\'n$^a$, B.~Kohlmeyer$^d$, F.~Laue$^{a,*}$, M.~Menzel$^d$,
L.~Naumann$^f$, E.~Schwab$^a$, P.~Senger$^a$, Y.~Shin$^c$,
H.~Str\"obele$^c$, C.~Sturm$^{b,a}$, G.~Sur\'owka$^{a,e}$,
F.~Uhlig$^{b,a}$,
A.~Wagner$^f$, W.~Walu\'s$^e$\\
KaoS Collaboration\\
and B.~K\"ampfer$^f$, H.W.~Barz$^f$\\
(KaoS Collaboration) \\
$^a$ Gesellschaft f\"ur Schwerionenforschung, D-64220 Darmstadt, Germany\\
$^b$ Technische Universit\"at Darmstadt, D-64289 Darmstadt, Germany\\
$^c$ Johann Wolfgang Goethe-Universit\"at, D-60325 Frankfurt am Main, Germany\\
$^d$ Phillips Universit\"at, D-35037  Marburg, Germany\\
$^e$ Jagiellonian University, PL-30059 Krak\'ow, Poland\\
$^f$ Forschungszentrum Rossendorf, D-01314 Dresden, Germany \\
$^g$ Technische Universit\"at Dresden, D-01062 Dresden, Germany\\
$^{\dagger}$ Present address: CERN, CH-1211 Geneve 23,
Switzerland\\
$^*$ Present address: Brookhaven National Laboratory,  USA}

\date{\today}

\begin{abstract}
  Differential production cross sections of $K^{\pm}$
  mesons have been measured in $p$ + C and $p$ + Au collisions at
  1.6, 2.5 and 3.5 GeV proton beam energy. At beam energies close to
  the production threshold, the $K^-$ multiplicity is
  strongly enhanced with respect to proton-proton collisions.
  According to microscopic transport calculations, this enhancement is
  caused by two effects: the strangeness exchange reaction
  $NY \to K^- NN$ and an attractive in-medium $K^-N$
  potential at saturation density.
\end{abstract}

\maketitle

The last decade witnessed substantial experimental and theoretical
efforts in the study of in-medium properties of strange particles.
In particular, a large body of new data on the production of kaons
and antikaons in nucleus-nucleus collisions at beam energies below
or close to the nucleon-nucleon ($NN$) threshold has been
collected. It was found that the $K^-$/$K^+$ ratio is enhanced in
heavy-ion collisions as compared to proton-proton collisions
\cite{barth,laue,menzel,foerster}. The enhanced production of
$K^-$ mesons per number of participants was found to be partly due
to strangeness exchange reactions ($\pi  Y \to K^- N$ with $Y =
\Lambda,\Sigma$) which are strongly suppressed in $p+p$ reactions.
Nevertheless, the measured kaon and antikaon yields can only be
reproduced by transport model calculations when taking into
account density-dependent $K$ meson nucleon ($KN$) potentials
parameterizing effectively the in-medium modification of the $K$
mesons \cite{cass_brat,li_brown}. The pronounced patterns of the
elliptic and directed flow of $K^+$ mesons provide independent
hints for the existence of a repulsive kaon-nucleon in-medium
potential \cite{shin,crochet}.

The quantitative study of in-medium effects is complicated by the
fact that the KN potentials  depend on the nuclear density which
varies strongly with time during the course of a nucleus-nucleus
collision. In order to avoid this complication, one can
investigate proton-nucleus collisions where the nuclear density is
well defined during particle production. Already at saturation
density the in-medium effects are expected to influence
strangeness production  at beam energies close to the production
threshold in $NN$ collisions (1.58 GeV for $ p p \to K^+ \Lambda
p$ and 2.5 GeV for $ p p \to K^+ K^- p p$). Whereas some data
exist on $K^+$ meson production in proton-nucleus collisions at
beam energies between 1.2 and 2.5 GeV proton energy
\cite{debowski,anke}, very little is known about $K^-$ production
in proton-nucleus collisions at threshold beam energies.
Calculations predict a measurable in-medium effect on the yield
and the phase space distributions of antikaons produced in
proton-nucleus collisions at threshold
energies~\cite{Sibirtsev:1998vz}.

So far only two experiments on antikaon production in
proton-nucleus collisions close to threshold energies have been
performed. At the KEK-PS $K^-$ mesons were measured at a fixed
laboratory angle of 5.1 degrees and at a fixed $K^-$ momentum of
1.5 GeV/c using proton kinetic energies of 3.5, 4.0, and 5.0 GeV
and light targets (C, Cu) \cite{kek}. At the ITEP-PS $K^-$ mesons
were measured at a fixed laboratory angle of 10.5 degrees and at a
fixed $K^-$ momentum of 1.28 GeV/c using proton kinetic energies
between 2.3 and 2.92 GeV and light targets (Be, Al) \cite{itep}.
In both cases, however, neither the emission angle nor the
momentum of the antikaons have been varied. These data are
difficult to interpret as the extrapolation of the measured
antikaon yields to full phase space are affected by uncertainties
in the momentum distribution of the antikaons and their angular
distribution.

In order to improve the data situation and to disentangle the
above mentioned effects  we performed a systematic investigation
of $K^+$ and $K^-$ meson production  in proton-nucleus collisions.
In this Letter we present results of experiments using proton
beams of 1.6, 2.5 and 3.5 GeV kinetic energy impinging on a light
and a heavy target. For the first time the spectral and angular
distributions of antikaons have been measured in this energy
regime.

The experiments were performed with the Kaon Spectrometer, KaoS,
at the heavy-ion synchrotron, SIS,  at GSI in Darmstadt
\cite{senger}. We bombarded C and Au targets (thickness 7 mm and 2
mm, respectively) with a proton beam having an intensity of up to
$10^{11}$ protons per spill (spill duration $\sim$ 10 s). The
emitted charged particles were detected at laboratory angles of
$\theta_{\rm lab}$ = 40$^{\circ}$, 48$^{\circ}$, 56$^{\circ}$, and
64$^{\circ}$. By using three settings of the magnetic field a
coverage of laboratory momenta from $p_{\rm lab}$ = 0.3 GeV/c to
1.1 GeV/c was obtained.

\begin{figure}
  \centering \includegraphics[width=0.65\textwidth]{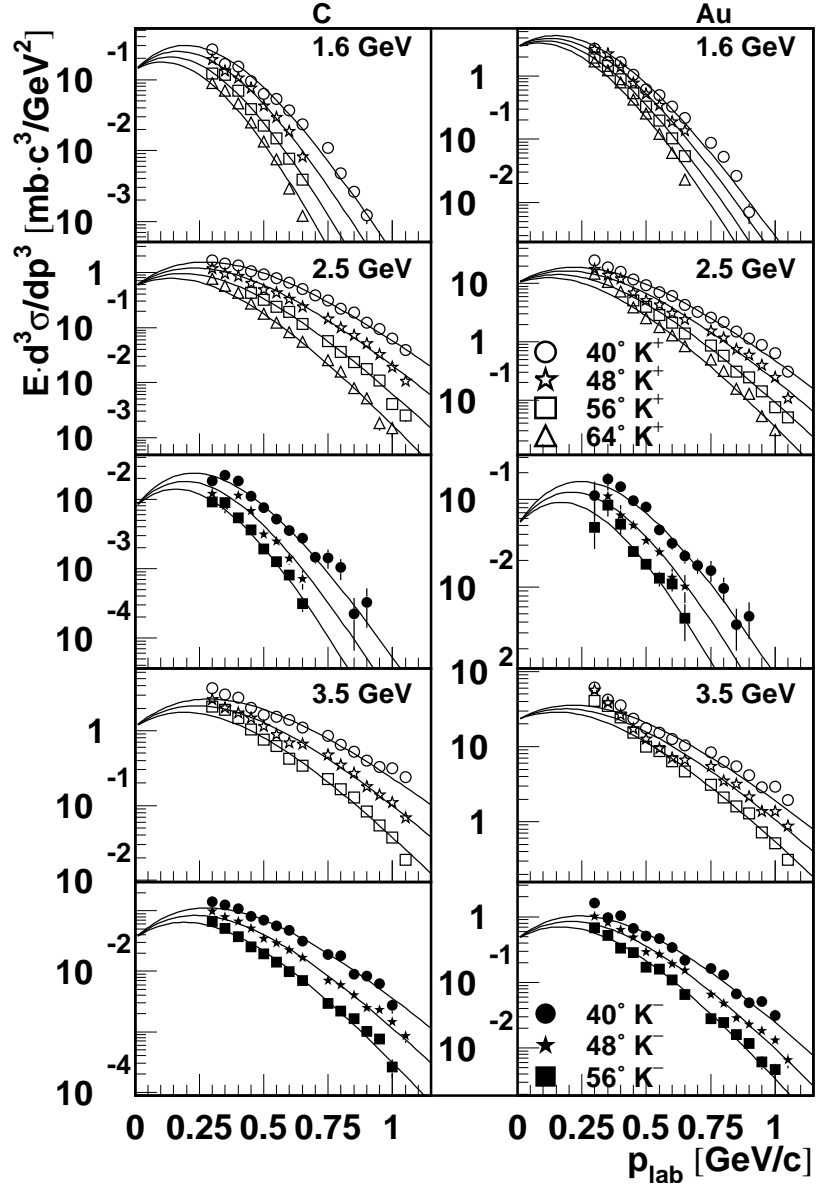}
  \caption{Invariant production cross sections of $K^+$ (open symbols)
    and $K^-$ mesons (full symbols) for inclusive proton-carbon (left
    row) and proton-gold collisions (right row) at 1.6, 2.5, and 3.5
    GeV (from top to bottom) as a function of laboratory momentum. The
    lines correspond to the distribution~(\ref{eq:juttner}) fitted to
    the data (see text and Table~\ref{Table1}).}
\label{labspec}
\end{figure}

Figure~\ref{labspec} shows the production cross sections for $K^+$
and $K^-$ mesons  as a function of their laboratory momentum
measured in inclusive proton-carbon (left column) and proton-gold
collisions (right column) at 1.6, 2.5, and 3.5 GeV proton energy
(from top to bottom). The laboratory angles are indicated. The
error bars represent the statistical uncertainties. An overall
systematic error of 15\% due to efficiency corrections and beam
normalization has to be added.

The lines in Figure~\ref{labspec} represent a
J\"uttner like distribution function
\begin{equation}
E \frac{d \sigma}{d^3 p} = \frac{\sigma_{\mathrm{fit}}}{4 \pi m_K^2 T K_2 (m_K /T)} (p \cdot u)
\exp \left(- \frac{(p \cdot u)}{T} \right)
\label{eq:juttner}
\end{equation}
with $p \cdot u \equiv (E_{\rm lab} - \beta p_{\rm lab} \cos
\theta_{\rm lab}) / \sqrt{1 - \beta^2}$, where $E_{\rm lab}$ is
the $K^\pm$ energy in the laboratory system and $\beta$ is the
velocity of the kaon emitting source; $m_K$ denotes the $K^\pm$
rest mass, and $K_2$ stands for a modified Bessel function. In a
local rest   frame ($\beta = 0$), the J\"uttner distribution
reduces to the known   Maxwell-Boltzmann distribution function.
Expression (\ref{eq:juttner}) was fitted simultaneously to the set
of differential cross sections measured at different laboratory
angles for each system at each energy. In this approach it is
assumed that the particles were emitted isotropically from a
source which moves with the velocity $\beta$ and which has a
momentum  distribution characterized by a temperature $T$. The
results of this fitting procedure for the source velocity, the
temperature and the total production cross section
$\sigma_{\mathrm{fit}}$ are listed in Table~\ref{Table1}. The
errors of the fit parameters $\beta$ and $T$ are always on the
order of $3-5\%$. The production cross section
$\sigma_{\mathrm{fit}}$ is affected by an additional systematic
error of $20\%$ due to uncertainties in the relative normalization
of the measurements performed at different magnetic field values.
This error has to be added to the values given in
Table~\ref{Table1}.

The values of $\beta$ for proton-nucleus collisions are substantially
smaller than the corresponding center-of-mass velocity of the
nucleon-nucleon system. This slowing-down of the apparent source
velocity may be caused by two effects: (i) the impinging proton
collides with a cluster of 2 - 4 correlated target nucleons, and (ii)
the kaons are back-scattered at the target nucleus. In both cases, the
source velocity $\beta$ should be larger for proton-carbon than for
proton-gold collisions. Indeed, this is observed for $K^+$ mesons.  In
the case of $K^-$ mesons, however, the velocity $\beta$ is almost
independent of the target nucleus mass. This observation indicates
that the production process of $K^-$ mesons is more involved.

The apparent temperature $T$ is significantly smaller for $K^-$ mesons
than for $K^+$ for the same collision system. A similar observation
was made in nucleus-nucleus collisions at near-threshold beam
energies~\cite{foerster}. The interpretation of this effect was
related to the delayed emission of $K^-$ mesons due to strangeness
exchange reactions like $\pi Y \to K^- N$ with $Y = \Lambda,\Sigma$.

\begin{figure}
  \centering \includegraphics[width=14cm]{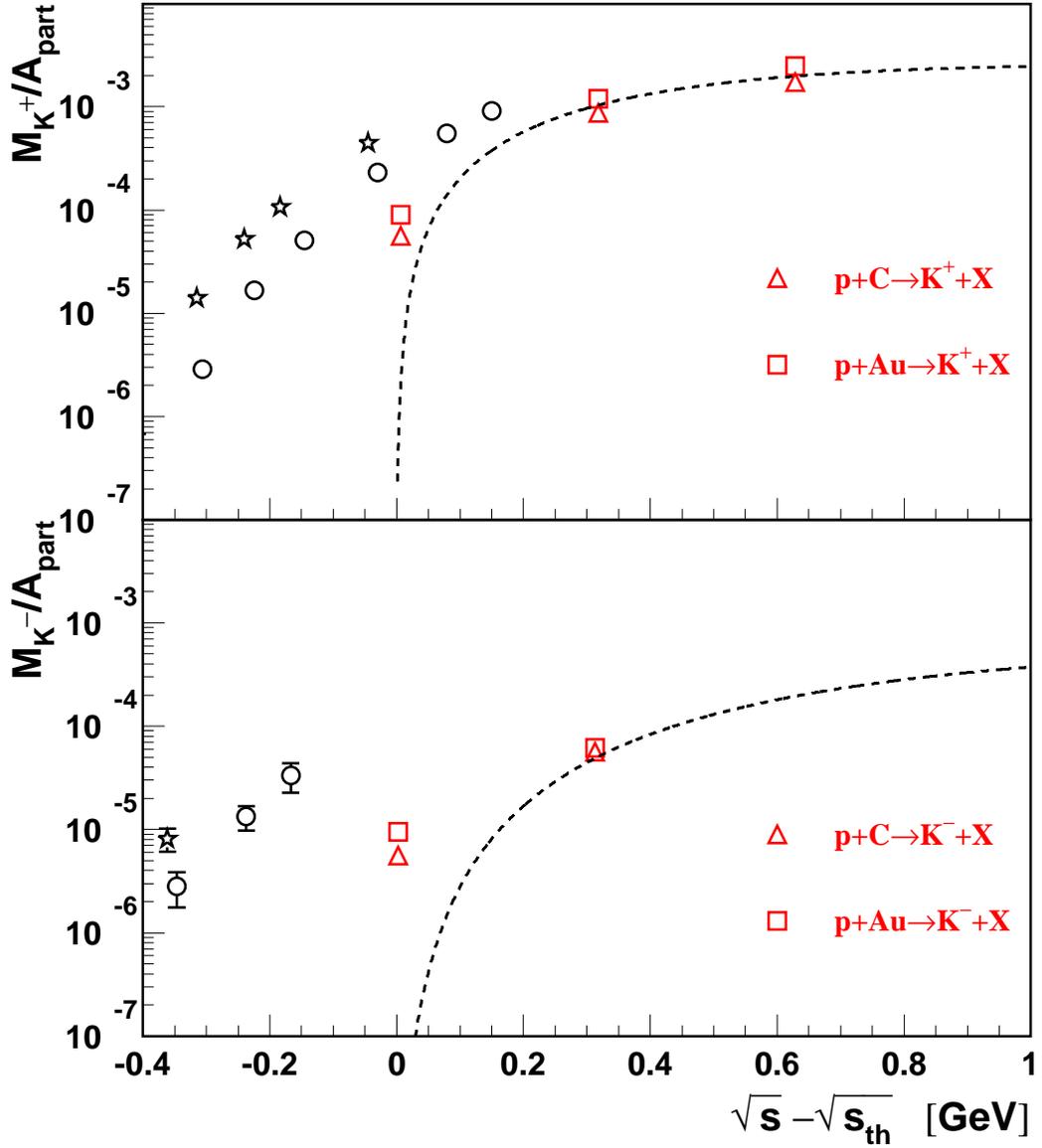}
  \caption{Multiplicities of $K^+$ mesons (upper panel) and $K^-$
    mesons (lower panel) per participating nucleon for proton-carbon (triangles) and proton-gold
    collisions (squares) as a function of the available energy in the
    $NN$ center-of-mass system (i.e.~the Q-value). The dashed curves
    correspond to parameterizations \cite{sibirtsev} of the measured $p+p$
    data. Data for carbon-carbon (circles) and gold-gold
    (stars) are taken from~\cite{laue} and ~\cite{sturm,foerster}, respectively. }\label{multi}
\end{figure}

In order to visualize the effect of the nuclear medium on
strangeness production we compare our data  from proton-nucleus
collisions to  data measured in proton-proton and nucleus-nucleus
collisions.  Figure~\ref{multi} shows the $K^+$ and $K^-$
multiplicities $M_K$ normalized to the number of participating
nucleons $A_{\mathrm{part}}$ as a function of the excess energy.
The multiplicity is defined as $M_K =
\sigma_{\mathrm{fit}}/\sigma_R$ with $\sigma_R = \pi R^2$ the
geometrical reaction cross section. Using $R = (0.6 + 1.2\times
A^{1/3})\,\mathrm{fm}$ (proton radius 0.6 fm) one obtains
$\sigma_R$ = 0.35 b for $p+$C and $\sigma_R$ = 1.8 b for $p+$Au.
The excess energy is defined as the difference between the energy
available in a free $NN$ collision and the $K^+$ or $K^-$
production threshold energies in $NN$ scattering. For p+A
collisions, we define $A_{\mathrm{part}}$ via the source velocity
$\beta = p_{\mathrm{beam}}/(E_{1\mathrm{beam}}+m_2)$. Here
$p_{\mathrm{beam}}$ and $E_{1beam}$ denote the beam proton
momentum   and total energy and $m_2$ is the mass of the target
nucleons  involved in the $K^{\pm}$ production. Hence,  for p+A
collisions we  obtain $A_{\mathrm{part}} = 1 + m_2/m_N$ with the
nucleon mass  $m_N$. For impact parameter integrated A+A
collisions, the average number of participating nucleons equals
A/2, while for proton-proton collisions we take
$A_{\mathrm{part}}=2$.

At beam energies well above threshold, the values of
$M_{K^{\pm}}/A_{\mathrm{part}}$ from proton-nucleus collisions
agree with the ones from proton-proton collisions (represented by
the dashed lines). At threshold energies, however, the
proton-nucleus data clearly exceed the proton-proton data, but
undershoot significantly the nucleus-nucleus data. The enhancement
of $M_{K^{\pm}}/A_{\mathrm{part}}$ when going from p+p, over p+A
to A+A collisions at threshold beam energies is mainly due to
Fermi motion, secondary collisions, and increased density. These
effects lower the effective thresholds and cause a large
difference in $M_{K^{\pm}}/A_{\mathrm{part}}$ between p+p and p+A,
but only a moderate difference between p+C and p+Au, where the
number of participant nucleons differ less than a factor of 2, and
where the density is comparable. Multiple collisions involving
several projectile nucleons can only occur in A+A but not in p+A
reactions. In addition, multiple collisions occur more frequently
in A+A  than in p+A due to the increased density. These effects
are nonlinear in $A_{\mathrm{part}}$, and lead to an enhancement
of $M_{K^{\pm}}/A_{\mathrm{part}}$ in  A+A collisions with respect
to p+A collisions.

The data presented in figure~\ref{multi} indicate that the
enhancement factor between p+p, p+A, and A+A collisions is
significantly larger for $K^-$ mesons than for $K^+$ mesons. Such
an  effect is expected for an increasing contribution of
strangeness exchange reactions to $K^-$ production, and for
density dependent in-medium effects.  Strangeness exchange
reactions are also responsible for the low apparent temperature of
the $K^-$ meson spectra as compared to the $K^+$ meson spectra,
and for the weak dependence of the source velocity $\beta$ on the
target mass: The $K^-$ mesons freeze-out at a late stage of the
collision when the system has lost its memory on entrance channel
effects.

It is interesting to note that the difference in
$M_{K^-}/A_{\mathrm{part}}$ between p+C and p+Au collisions is
relatively small, similar to the $K^+$ mesons. This observation
again indicates the important role of strangeness exchange
reactions: The yield of $K^-$ mesons created via strangeness
exchange depends only on the abundance of hyperons, and not on the
numbers of nucleons in the system, as both $K^-$ production ( via
$N Y \to K^- N N$) and $K^-$ absorption (via $K^- N \to Y \pi$ )
is proportional to the number of nucleons. Therefore, the target
mass dependence of $K^-$ and $K^+$ production is very similar.
Moreover, the in-medium potentials depend not on mass, but on the
nucleon density which does not differ much between p+C and p+Au
collisions.

\begin{figure}
\includegraphics[width=14cm]{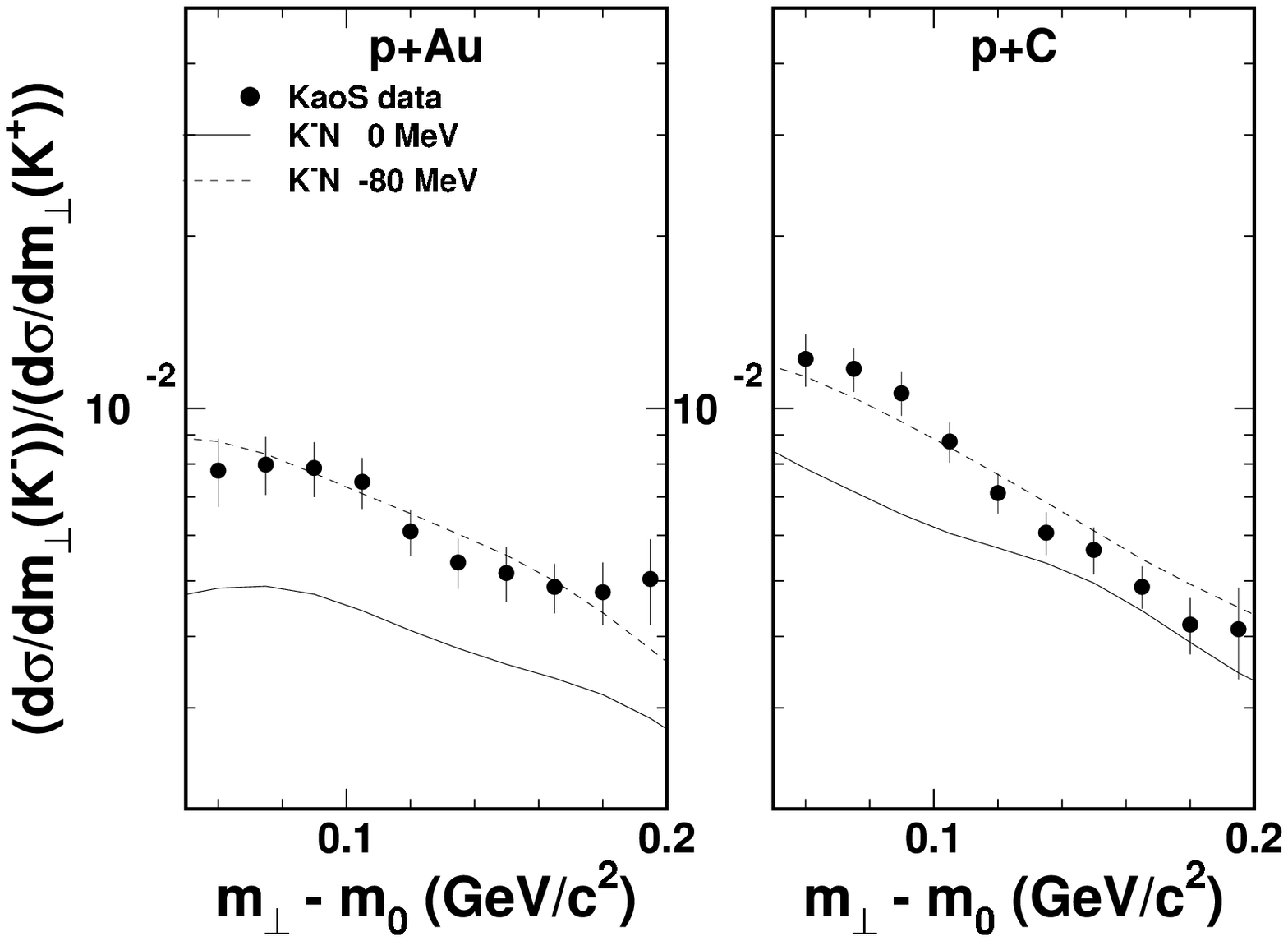}
\caption{Ratio of invariant production cross sections of $K^-$ mesons
  over $K^+$ mesons for inclusive proton-gold (left panel) and
  proton-carbon collisions (right panel) as a function of transverse
  mass. The data (full circles) were taken at a beam energy of 2.5 GeV
  and were integrated over laboratory angles between
  $\theta_{\mathrm{lab}} = 36^{\circ}$ and $60^{\circ}$. The solid and
  dashed curves depict results of BUU transport model calculations
  including strangeness exchange as well as a $K^+N$ potential of
  $+25$ MeV. These calculations use $K^-N$ potentials as indicated.}
\label{BUU}
\end{figure}

The interplay of production and absorption of $K^-$ mesons
via strangeness exchange reactions depends on the in-medium
properties of the strange particles. For example, a $K^-N$
potential in the nuclear medium can modify the Q-values for these
reactions. In order to study the role of in-medium effects on
$K^-$ production in $p$+A collisions more quantitatively, we
discuss the ratio
\begin{equation}
R = \left.\frac{d\sigma}{dm_{\perp}}\right|_{K^-}\Biggl.\Biggr/\left.\frac{d\sigma}{dm_{\perp}}\right|_{K^+}
\label{eq:1}
\end{equation}
of invariant cross sections of inclusive $K^-$ over $K^+$
production as a function of the transverse mass in $p$+C and
$p$+Au collisions at an energy of 2.5 GeV (see Figure~\ref{BUU}).
The transverse mass $m_{\perp}$ is defined as $\sqrt{p_{\perp}^2 +
m_K^2}$ with $p_{\perp}$ the transverse momentum. The data
presented in Figure~\ref{BUU} are integrated over the measured
angular range (from 36$^{\circ}$ to 60$^{\circ}$) and compared to
results of BUU calculations. A description of the BUU model
employed may be found in \cite{Barz:2003wz}.  The BUU calculations
represented by the solid and dashed lines take into account the
strangeness exchange reactions $\pi Y \to K^- N$ and $N Y \to K^-
NN$. The first channel is dominant in Au+Au collisions. In the C+C
system, both channels are about equally important.  In
proton-nucleus collisions, however, it turns out that the process
$N Y \to K^- NN$ contributes about twice as much to $K^-$
production as the $\pi Y\to K^- N$ reaction. The reason for that
is the low probability to create both a pion and a $\Lambda$
hyperon with only one proton as the projectile. According to the
BUU calculations,  the strangeness exchange reactions dominate
over the direct $K^{-}$ production processes, both in
nucleus-nucleus as well as proton-nucleus collisions. For example,
in proton-nucleus at 2.5 GeV the contribution of strangeness
transfer processes amounts to 50-60\%, and for nucleus-nucleus
collisions it is even larger (70-80\%).

In order to discuss the possible role of density dependent
in-medium potentials, it is important to know at which densities
the finally observed $K^-$ mesons are produced in the various
collision systems. For example, for $p$+C ($p$+Au) collisions at a
beam energy of 2.5 GeV the BUU calculations predict an average
density about 0.8 (0.9) times the normal nuclear matter density,
whereas for C+C (Au+Au) collisions the average value is slightly
above 1.1 (1.5) times saturation density. The observed $K^+$
mesons are produced earlier, and, hence, at larger average
densities: the BUU calculations predict a value of 0.9 (1.0) times
normal nuclear matter density for $p$+C ($p$+Au) collisions, while
for C+C (Au+Au) collisions a value of about 1.4 (2.2) is obtained
for the same beam energy per nucleon.

Figure~\ref{BUU} depicts the $K^-/K^+$ ratio as function of
transverse mass for p+Au (left panel) and p+C collisions (right
panel) at an proton beam energy of 2.5 GeV. The data (full dots)
are compared to the results of BUU calculations with and without
in-medium $K^-N$ potentials. The theoretical uncertainties in the
$K^+$ production channels -- which directly affect the $K^-$ meson
yield via the strangeness exchange reactions -- essentially cancel
out in the $K^-/K^+$ ratio. When assuming a $K^-N$ potential of
about $-80 $ MeV (dashed curve in Fig.~\ref{BUU}) the calculations
agree reasonably well with the $p+$Au data. In the case of $p$+C
collisions, the calculations slightly underestimate the low energy
part and overshoot the high energy part of the ratio. The
discrepancy between the data and the calculations neglecting
$K^-N$ potentials (solid lines) clearly demonstrates the important
role of in-medium effects. The calculations also take into account
a $K^+N$ potential of +25 MeV at saturation density. Calculations
based on momentum dependent potentials as discussed
in~\cite{Sibirtsev:1998vz} yield a similar agreement with the data
for an momentum averaged $K^-N$ potential of -80 MeV. From the
analysis of nucleus-nucleus data a value of -110$\pm$15 MeV has
been obtained for the  $K^-N$ potential at saturation density
\cite{cass_brat} (see also the compilation in~\cite{kolomeitsev}).
A more detailed analysis of our data with respect to in-medium
effects calls for an improved theoretical approach, e.g. transport
calculations including off-shell effects and in-medium spectral
functions \cite{Leupold:1999ga,Cassing:2003vz,tolos,lutz}.

In summary, we have presented experimental data on $K^+$ and $K^-$
production in $p+$C and $p+$Au collisions at beam energies close
to the production thresholds. This is the first measurement of
phase space distributions of antikaons in proton induced reactions
on nuclei in this energy range. The results presented for
$K^{\pm}$ production represent an important data set filling the
gap between nucleon-nucleon and nucleus-nucleus collisions. The
comparison of the data to results of microscopic transport
calculations indicate that the in-medium $K^-N$ potential is on
the order of -80 MeV at normal nuclear  density.

This work was supported by the German Federal Government (BMBF),
by the Polish Committee of Scientific Research (No. 2P3B11515) and
by the GSI fund for Universities.

\newpage

\begin{table*} [hbt]
  \caption{Beam energy, target, K meson
    species, production cross section, source temperature, and source
    velocity. The values of
    $\sigma_{\mathrm{fit}}$, $T$, and $\beta$ are obtained by the
    fitting procedure described in the text. A systematic error of 20\%
    has to be added to the values of $\sigma_{\mathrm{fit}}$. The center-of-mass
    velocity for the nucleon-nucleon system is $\beta$ = 0.68, 0.76,
    and 0.81 for beam energies of 1.6, 2.5, and 3.5 GeV,
    respectively.}\label{Table1}
\begin{ruledtabular}
\begin{tabular}{ccccccc}
E\ensuremath{_{\mathrm{beam}}} (GeV) & target & $K$ &
\ensuremath{\sigma_{\mathrm{fit}} }(mb) & $T$ (MeV)&
\ensuremath{\beta}  \\ \hline 1.6  & C   & \ensuremath{K^+}  &
0.070 \ensuremath{\pm}  0.006  &
45.6 &   0.48 \\
2.5  & C   & \ensuremath{K^+}  &   0.89 \ensuremath{\pm}   0.08   &
69.9 &   0.63 \\
3.5  & C   & \ensuremath{K^+}  &   2.04 \ensuremath{\pm}   0.25   &
88.2 &   0.65 \\
1.6  & Au  & \ensuremath{K^+}  &   0.83 \ensuremath{\pm}   0.05   &
49.0 &   0.37 \\
2.5  & Au  & \ensuremath{K^+}  &   8.53 \ensuremath{\pm}   0.95   &
73.0 &   0.53 \\
3.5  & Au  & \ensuremath{K^+}  &  20.74 \ensuremath{\pm}   1.69   &
90.8 &   0.52 \\
2.5  & C   & \ensuremath{K^-}  &   0.0069 \ensuremath{\pm} 0.0006 &
45.1 &   0.55 \\
3.5  & C   & \ensuremath{K^-}  &   0.065 \ensuremath{\pm}  0.007 &
65.9 &   0.64 \\
2.5  & Au  & \ensuremath{K^-}  &   0.059 \ensuremath{\pm}  0.008 &
51.1 &   0.58 \\
3.5  & Au  & \ensuremath{K^-}  &   0.48 \ensuremath{\pm}   0.06   &
68.1 &   0.58 \\
\end{tabular}
\end{ruledtabular}
\end{table*}

\end{document}